\documentclass[twocolumn, amssymb, amsmath, aps, prb, showpacs, 10pt]{revtex4-1}
\usepackage[dvips]{graphicx}
\usepackage[dvips]{color}
\begin{document}
\title{Spin glass like ground state and observation of exchange bias in Mn$_{0.8}$Fe$_{0.2}$NiGe alloy}
\author{P. Dutta$^1$}
\author{S. Pramanick$^2$}
\author{D. Venkateshwarlu$^3$}
\author{V. Ganesan$^3$}
\author{S. Majumdar$^2$}
\author{D. Das$^1$}
\author{S. Chatterjee$^1$}
\email{souvik@alpha.iuc.res.in}
\affiliation{$^1$UGC-DAE Consortium for Scientific Research, Kolkata Centre, Sector III, LB-8, Salt Lake, Kolkata 700 098, India}
\affiliation{$^2$Department of Solid State Physics, Indian Association for the Cultivation of Science, 2 A \& B Raja S. C. Mullick Road, Jadavpur, Kolkata 700 032, India}
\affiliation{$^3$UGC-DAE Consortium for Scientific Research,University Campus, Khandwa Road, Indore 452 017, India}
\pacs{75.50.Lk, 75.47.Np, 81.30.Kf}
\begin{abstract}
The ground-state magnetic properties of hexagonal equiatomic alloy of nominal composition Mn$_{0.8}$Fe$_{0.2}$NiGe were investigated through dc magnetization and heat capacity measurements. The alloy undergoes first order martensitic transition below 140 K with simultaneous development of long range ferromagnetic ordering from the high temperature paramagnetic phase. The undoped compound MnNiGe has an antiferromagnetic ground state and it shows martensitic like structural instability well above room temperature. Fe doping at the Mn site  not only brings down the martensitic transition temperature, it also induces ferromagnetism in the sample. Our study brings out two important aspects regarding the sample, namley (i) the observation of {\it exchange bias} at low temperature, and (ii) {\it spin glass} like ground state which prevails below the martensitic and magnetic transition points. In addition to the observed usual relaxation behavior the spin glass state is confirmed by zero field cooled memory experiment, thereby indicating cooperative freezing of spin and/or spin clusters rather than uncorrelated dynamics of superparamagnetic like spin clusters. We believe that doping disorder can give rise to some islands of antiferromagnetic clusters in the otherwise ferromagnetic background which can produce interfacial frustration and exchange pinning responsible for spin glass and exchange bias effect. A comparison is made with doped rare-earth manganites where similar phase separation can lead to glassy ground state.
\end{abstract}
\maketitle

\section{Introduction}
Intermetallic alloys and compounds, which undergo magneto-structural transition, are always in the forefront of research due to their rewarding properties such as large magnetoresistance~\cite{morellon-apl1,Algarabel-apl1,Havela-jap1}, large magnetocaloric effect~\cite{gs-mce1}, magnetic superelasticity~\cite{Kainuma-nat1,Krenke-prb1}, colossal magnetostriction~\cite{morellon-prb1} etc. Coupling between spin and lattice degrees of freedom plays pivotal role towards the observation of such magneto-functional properties. Magnetic equiatomic ternary alloys of general formula MM$^{\prime}$X (M, M$^{\prime}$ = transition metals, X = Si, Ge, Sn) belong to the series of compounds that undergo magneto-structural transition of martensitic type on cooling from high temperature.~\cite{zhang-apl}  These alloys are identified as new class of ferromagnetic shape memory alloys (FSMAs).~\cite{Koyama-fsma} Recent discovery of very large  magnetocaloric effect (MCE) around martensitic phase transition (MPT) triggered renewed interest on these ternary equiatomic alloys. Though several doping studies have been performed to explore the magnetocaloric properties of these materials,~\cite{ali-apl,Trung-apl1,Dincer-jalcom1,Caron-prb1} very little effort has been made to address the true nature of their magnetic ground state. 

\par
Stoichiometric MnNiGe equiatomic alloy is one of the well-studied systems among MM$^{\prime}$X series of alloys, which undergoes MPT at $T_t$ = 470 K during cooling and orders antiferromagnetically below $T_N$ = 346 K.~\cite{zhang-jpd,zhang-apl} These characteristic transition temperatures depend strongly on the composition of the alloy. One can control the structural transition temperature and induce ferromagnetism in this system by suitably doping different transition metals in the Mn or Ni site. Recently, Liu {\it et al.} have reported the effect of Fe doping at the Mn and Ni sites on the magneto-structural transition of this alloy.~\cite{liu-nc} Fe doping in MnNiGe alloy alters spiral antiferromagnetic (AFM) interaction among Mn atoms and induces ferromagnetic (FM) or spin glass (SG) like state in the sample. Strong dependence of magnetic and structural properties on Fe doping tempted us to investigate extensively the true nature of the magnetic ground state of this system of alloys. Here one such Fe-doped MnNiGe alloy of nominal composition Mn$_{0.8}$Fe$_{0.2}$NiGe has been investigated by different magnetic and heat capacity measurements.

\section{Experimental Details}
The polycrystalline sample Mn$_{0.8}$Fe$_{0.2}$NiGe was prepared by argon arc-melting and subsequent annealing at 800$^{\circ}$C for 100 h followed by a rapid quenching in ice water. Room temperature x-ray powder diffraction (XRD) pattern (see inset of fig.~\ref{mt}(a)) of the sample recorded using the Cu K$_{\alpha}$ radiation confirms single phase hexagonal Ni$_2$In-type structure (space group $p6_3/mmc$). The lattice parameters   are found to be $a$ = 4.077 \AA, $c$ = 5.320 \AA. The dc magnetization ($M$) of the sample was measured using a Quantum Design SQUID magnetometer (MPMS 7, Evercool model). Whereas heat capacity ($C_p$) of the sample was measured in a commercial Physical Properties Measurement System from Quantum Design using the relaxation technique.
 
\section{Results}

\begin{figure}[t]
\centering
\includegraphics[width = 7.5 cm]{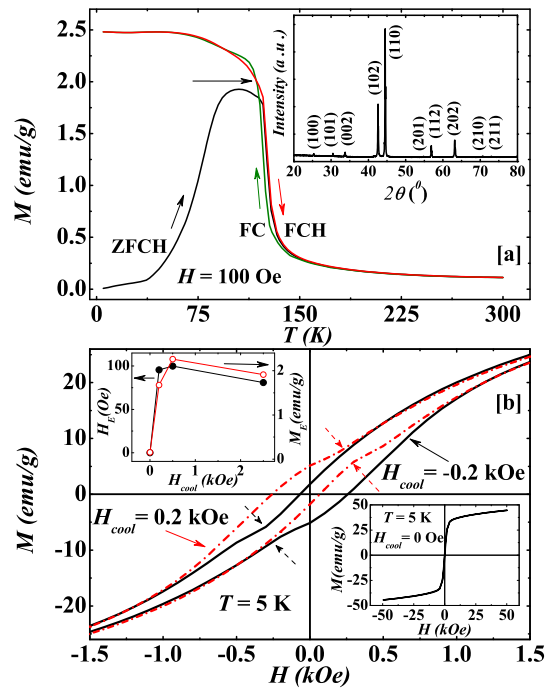}
\caption{(Color online) The main panel of (a) represents temperature dependence of magnetization ($M$) in presence of 100 Oe of applied magnetic field ($H$) in the zero-field-cooled heating (ZFCH), field-cooled (FC) and field-cooled heating (FCH) protocols. Inset of (a) shows powder x-ray diffraction pattern of Mn$_{0.8}$Fe$_{0.2}$NiGe at room temperature. Main panel of (b) depicts isothermal $M$ as a function of applied $H$ at 5 K after the sample being cooled in $H$ = 0.2, and -0.2 kOe from 300 K. A full $\pm$50 kOe loop on the zero-field-cooled state is shown in the lower inset of (b). Upper inset of (b) shows variation of exchange field ($H_E$) and shift in magnetization ($M_E$) as a function of cooling field ($H_{cool}$)}
\label{mt}
\end{figure}

The main panel of fig.~\ref{mt}(a) depicts the variation of $M$ as a function of temperature ($T$) in presence of 100 Oe of applied magnetic field ($H$) in zero field cooled heating (ZFCH), field cooling (FC) and field cooled heating (FCH) sequences. A sudden increase in $M(T)$ data is observed around 140 K with decreasing $T$ in all three sequences indicating the presence of magnetic and martensitic transition in the sample. Clear thermal hysteresis associated with $M(T)$ data around 140 K confirms the first order nature of the transition. The width of the hysteresis loop is quite narrow ($\sim$ 4 K), and a careful look reveals that the FC and FCH curves cross once within the hysteresis region (around 118 K and marked by an arrow in fig.~\ref{mt}(a)) constituting a double loop structure. Similar two or more consecutive thermal hysteresis loops are not uncommon among shape memory alloys, and the phenomenon is often attributed to multi-step martensetic process.~\cite{doubleloop-MT1} The ZFCH curve is found to deviate from the field cooled data below 125 K (just below MPT) and a broad peak is observed around 100 K.

\par
Fe doping at the Mn site favors ferromagnetism, while the undoped sample is antiferromagnetic. Therefore, in the partially Fe-doped sample, there is a possibility for the presence of both FM and AFM interactions. Existence of such two different magnetic phases may give rise to exchange bias effect (EBE). The EBE is defined as the shift in the center of $M(H)$ loop from the origin when the sample is field cooled from high $T$ (well above the AFM and FM transition points).~\cite{eb-rev1,eb-rev2} In general, EBE is observed in the materials having two different types of magnetic interfaces, namely FM/AFM, FM/SG and FM/ferrimagnet based multilayer etc. However, EBE is also observed in bulk polycrystalline materials without any well defined interface ~\cite{eb-rev1,eb-rev2}. We recorded isothermal $M(H)$ data at 5 K after cooling the sample from 300 K in presence of different applied $H$ (see main panel of fig.~\ref{mt}(b), $H_{cool}$ = 0.2, and - 0.2 kOe are shown for clarity). All the $M-H$ isotherms were recorded between $\pm$50 kOe, which is well above the technical saturation field of the sample and rules out the minor loop effect towards the observed EBE. A restricted range of $\pm$1.5 kOe is shown in the main panel of fig.~\ref{mt}(b) for clarity. A full loop between $\pm$50 kOe at 5 K for $H_{cool}$ = 0 kOe is shown in the lower inset of fig.~\ref{mt}(b) for completeness. Signature of EBE is clear from the shift observed in field-cooled $M-H$ isotherms both in horizontal ($H$) and vertical ($M$) axes. Shift in the horizontal field axis ($H_E$) and the vertical magnetization axis ($M_E$) in field cooled $M-H$ isotherm is the measure of EBE. Variations of $H_E$ and $M_E$ as a function of $H_{cool}$ are plotted in the upper inset of fig.~\ref{mt}(b). Both $H_E$ and $M_E$ increase sharply with $H_{cool}$ in the low field region and go through a maximum around 0.5 kOe followed by a sluggish drop with further increase of $H_{cool}$. For $H_{cool}$ = 500 Oe, $M_E$ is found to be 2.26 emu/g, which is close to the zero-field-cooled value of remnant magnetization $M_r$ = 1.99 emu/g. Initial increase in $H_{cool}$ (before the saturation of FM fraction) favors FM fraction which results sharp increase in EBE, whilst large value of $H_{cool}$ weakens the interfacial exchange coupling and hence the EBE decreases.

\par
In addition to EBE, both zero-field-cooled (not shown here) and field-cooled $M-H$ isotherms show a clear signature of constriction near the origin (marked by dotted arrows in main panel of fig.~\ref{mt}(b)). Although the reason behind such constricted double loop structure of hysteresis is not exactly clear, often complex domain structure of the AFM phase is held responsible for similar anomalies in hysteresis loop in magnetically inhomogeneous materials.~\cite{Dobrynin-njp}

\par
\begin{figure}[t]
\centering
\includegraphics[width = 7.0 cm]{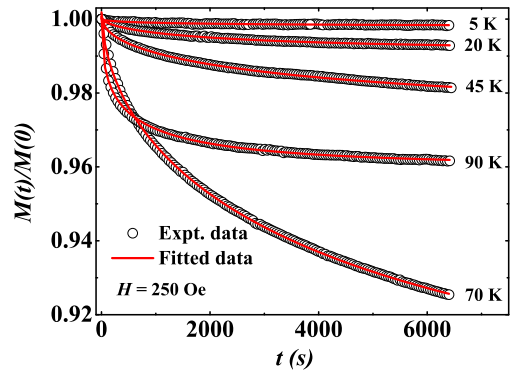}
\caption{(Color online) Isothermal variation of magnetization ($M$) as a function of time ($t$) at different constant temperatures in zero magnetic field after being cooled in presence of 250 Oe of applied $H$ from 300 K. Here the data have been normalized by the initial value of the magnetization [$M(0)$] at $t$ = 0. Solid lines are the modified stretched exponential fit to the experimental data.}
\label{rlx}
\end{figure}

Observation of EBE at 5 K indicates the presence of two types of magnetic phases in the system. To shed more light, we measured isothermal variation of $M$ as a function of time ($t$) at different constant temperatures (see fig.~\ref{rlx}, where $M$ is plotted in normalized form as $M(t)/M(0)$). The sample was first cooled from 300 K to the desired $T$ in presence of $H$ = 250 Oe and $M$ was measured as a function of $t$ immediately after removing $H$. The magnitude of relaxation is found to increase with increasing $T$ and attains a maximum at 70 K. On further increase in $T$, a drop in the relaxation magnitude is observed. After 6400 s of measurement, about 7.5\% change in $M$ is observed at 70 K. Such large change in $M$ indicates that the system evolves through a landscape of  spatially distributed free energy barriers. This relaxation behavior can be well explained by a modified stretched exponential (MSE) function given by $M(t) = M_0 + b \exp (-t/\tau)^\beta$. It is widely used to describe the relaxation behavior of glassy magnetic systems.~\cite{ito-prl1,rlx-prl2,rlx-rpp,mydosh-book,binder-rmp} Here $M_0$ corresponds to initial magnetization, $b$ is contribution from the glassy part, $\tau$ is the time constant and the exponent $0\leq\beta\leq 1$ is linked with the distribution of energy barriers among the metastable states. For an ordered FM system, the number of local energy minima shrink to a single global energy minimum and $\beta$ becomes unity. In our sample, the fitted values of $\beta$ for all $T$'s remain between 0.45 to 0.6 which fall within the range of $\beta$ values reported earlier for different SG-like systems.~\cite{wang-prb2,Chu-prl1} Apart from MSE model, we have also tried to fit the relaxation data with another slightly different function $M(t) = M_0 \exp(\frac{-C(t/\tau)^{(1-n)}}{(1-n)})$ used for glassy magnetic state.~\cite{Ralph,wu-prb} We find that the fitting to our data is relatively poor with the second function. However, the fitted values of  $n$ increase with increasing $T$ as expected for SG systems.~\cite{Ralph,wu-prb}

\par
\begin{figure}[t]
\centering
\includegraphics[width = 7.5 cm]{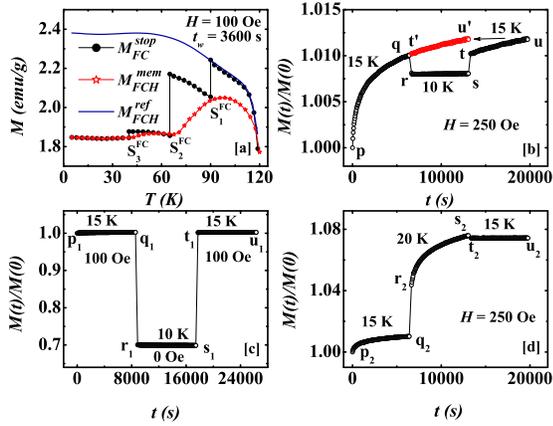}
\caption{(Color online) (a) Field-cooled field stop memory effect in dc magnetization {\it vs.} temperature data. The memory measurement was performed by cooling the sample in $H$ = 100 Oe with intermediate zero-field stops at three different temperatures (eg. S$^{FC}_1$ , S$^{FC}_2$ and S$^{FC}_3$, curve $M^{stop}_{FC}$) followed by uninterrupted heating in 100 Oe ($M^{mem}_{FCH}$). The reference curve $M^{ref}_{FCH}$ was also recorded for the sample on heating after being cooled in 100 Oe without intermediate stops. (b) Relaxation memory measurement in presence of 250 Oe of applied field. Time evolution of $M$ was measured with a small negative $T$ cycling in three different segments (eg. $\overline {pq}$ (15 K), $\overline {rs}$ (10 K) and $\overline {tu}$ (15 K)) of about 6400 s each. Here the $\overline {{t}^{\prime} {u}^\prime}$ segment is obtained by shifting $\overline {tu}$ to merge its starting point with the end point of $\overline {pq}$. (c) A similar relaxation memory measurement, with the only exception being that the intermediate 10 K data were recorded in zero field. (d) Rejuvenation (positive $T$ cycling) measurement of magnetic relaxation.}
\label{mem}
\end{figure}

In order to examine the presence of nonergodicity in the sample, we performed magnetic memory measurement using different protocols.~\cite{salamon-prl,nordblad-prl1} Fig.~\ref{mem}(a) represents field-cool field-stop memory measurement data. During this protocol, the sample was cooled in presence of 100 Oe of $H$ from room $T$ to the lowest $T$ of measurement (here 5 K) with three intermediate stops of 3600 s each at $T_{stop}$ = 90 K ($S_1^{FC}$), 65 K ($S_2^{FC}$) and 40 K ($S_3^{FC}$). During each isothermal stop, applied $H$ was reduced to zero. $H$ was reapplied during further cooling. This results step like drops in $M-T$ data (see curve $M^{stop}_{FC}$ in Fig.~\ref{mem}(a)). Subsequent heating in 100 Oe of applied $H$ (without any intermediate stop) shows characteristic feature at each $T_{stop}$, revealing the presence of magnetic memory in the present alloy (see curve $M^{mem}_{FCH}$ in Fig.~\ref{mem}(a)). A normal FCH curve ($M_{FCH}^{ref}$) in 100 Oe of applied $H$ (without any stop during cooling) is also plotted for reference.

\par
To strengthen the observed memory effect in $M-T$ data, we performed relaxation memory measurements with negative $T$ cycling as shown in fig.~\ref{mem} (b) and (c). During this measurement, relaxation data were recorded in zero field cooled (ZFC) mode. First the sample was cooled from room $T$ to 15 K in absence of external $H$ and subsequently $M$ was recorded as a function of time in presence of 250 Oe of applied $H$ for 6400 s ($\overline {pq}$). Then the sample was cooled down to 10 K in constant $H$ and $M(t)$ was recorded for another 6400 s ($\overline {rs}$) and finally, the sample was heated back to 15 K and $M(t)$ was recorded for further 6400 s ($\overline {tu}$). It is observed that $\overline {tu}$ part is simply a continuation of $\overline {pq}$ part, which reflects that the state of the sample before cooling is recovered when the sample is cycled back to initial $T$. To check the robustness of the relaxation memory effect, we performed both $T$ and $H$ cycling simultaneously(fig.~\ref{mem} (c)). The measurement protocol is same as depicted in fig.~\ref{mem} (b). The only difference is that during 10 K relaxation measurement, $H$ (100 Oe in this case) was reduced to zero. The continuity of $\overline {p_1q_1}$ and $\overline {t_1u_1}$ segments of the $M(t)$ curve before and after $T$ and $H$ cycling indicates that the sample remembers its previous state even after experiencing large change in $M$. For a non-ergodic system, a small positive $T$ cycling can erase the previous memory and rejuvenates the system. Such rejuvenation measurement was also performed and depicted in fig.~\ref{mem} (d). During this measurement, the sample was heated to a higher value of $T$ for the intermediate relaxation measurement, instead of a negative $T$ cycling as done in the relaxation memory measurement. Here $M$ {\it vs.} $t$ was measured at 15 K ($\overline {p_2q_2}$), 20 K ($\overline {r_2s_2}$), and 15 K ($\overline {t_2u_2}$) respectively. The result of this measurement is found to be significantly different (not a continuation of similar $M$) from the relaxation memory data described in fig.~\ref{mem} (b) and (c). This observation indicates that a small positive $T$ cycling can destroy the magnetic memory of the compound.

\par
\begin{figure}[t]
\centering
\includegraphics[width = 7.0 cm]{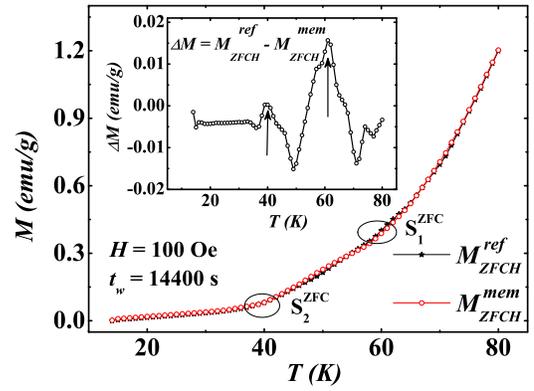}
\caption{(Color online) Main panel shows memory measurement in zero field cooled condition in dc magnetization {\it vs.} temperature data. The sample was first cooled in absence of external field with two intermediate stops of 14400 s each (eg. S$^{ZFC}_1$ and S$^{ZFC}_2$). Sample was then reheated in presence of 100 Oe applied $H$ up to 80 K ($M^{mem}_{ZFCH}$)). Zero field cooled heating reference curve ($M^{ref}_{ZFCH}$) with no intermediate stop during cooling was also recorded. The difference in magnetization $\Delta M = M^{ref}_{ZFCH} - M^{mem}_{ZFCH}$ is plotted as a function of $T$ in the inset.}
\label{zfcm}
\end{figure}

Signature of magnetic memory in the FC magnetization and relaxation data are a convincing evidences for the glassy magnetic state. However, such memory effect can also originate from the distribution of magnetic relaxation time among noninteracting superparamagnetic nanoparticles.~\cite{dattagupta-memory,nordblad-memory} In a phase separated system like Mn$_{0.8}$Fe$_{0.2}$NiGe, a similar effect may also arise from the independent relaxation of metastable phase clusters. To rule out this possibility, we performed zero field cooled (ZFC) memory measurement in $M$ {\it vs.} $T$ data (see main panel of fig.~\ref{zfcm}).~\cite{nordblad-memory} In this protocol, the sample was first cooled from 300 K to 5 K in absence of $H$ with two intermediate stops of 14400 s each ({\it e.g.} point S$^{ZFC}_1$ and point S$^{ZFC}_2$ in main panel of fig.~\ref{zfcm}). The sample was then heated back to 80 K in presence of 100 Oe of $H$. Characteristic features are observed around the stoppage points. These features are clearer in the difference curve $\Delta M = M^{ref}_{ZFCH} - M^{mem}_{ZFCH}$ (see inset of fig.~\ref{zfcm}). Since ZFC memory is recorded after cooling the sample in zero field, the positive signature of memory can not be attributed to the distribution of relaxation time for independent superparamagnetic clusters.~\cite{nordblad-prl1,dattagupta-memory,nordblad-memory} Therefore it confirms that the metestability in Mn$_{0.8}$Fe$_{0.2}$NiGe is due to spin glass like ground state and not connected to superparamagnetism. It is to be noted that aging effect measurement of $M$ is also traditionally performed to identify an SG ground state. However, there are several reports which show that aging effect can also be observed in superparamagnetic materials,~\cite{nordblad-prl1,sasaki-prb1} and therefore, we have not provided aging data to strengthen our observation.

\par
\begin{figure}[t]
\centering
\includegraphics[width = 7.5 cm]{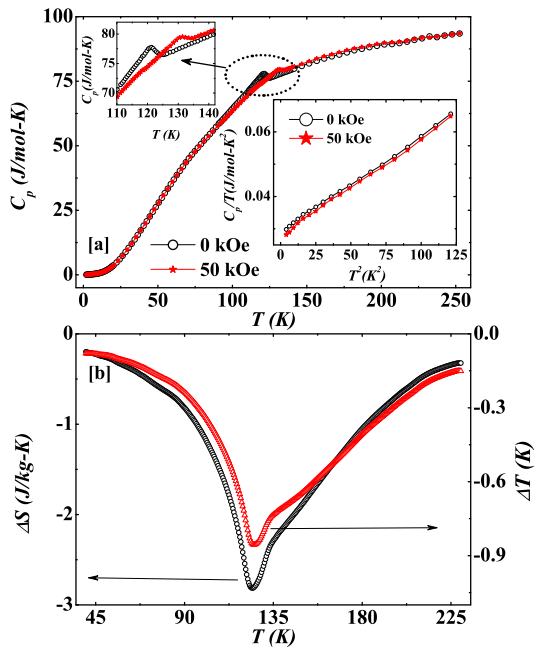}
\caption{(Color online) Main panel of (a) depicts temperature variation of heat capacity ($C_p$) in presence of zero and 50 kOe of external magnetic field during heating. Upper inset of (a) shows a zoomed version of the transition region. $C_p/T$ vs. $T^2$ is plotted in the lower inset of (a). $T$ variation of entropy change ($\Delta S$) and temperature change ($\Delta T$) are plotted in (b).}
\label{cp}
\end{figure}

We recorded $T$ variation of heat capacity ($C_p$) during heating from low-$T$ in zero field as well as in presence of 50 kOe of applied $H$ (see main panel of fig.~\ref{cp} (a)). Clear signature of MPT is observed around 125 K, which matches well with the $M-T$ data. Transition temperature is found to be shifted towards higher $T$ in presence of external $H$. Non linear nature of $C_p/T$ vs. $T^2$ data plotted in the lower inset of fig.~\ref{cp}(a) indicates that in addition to electronic and lattice contributions, some additional magnetic contribution is also present in the sample. To analyze the electronic, magnetic and lattice contributions in low-$T$ part (below 12 K) of $C_p$ vs $T$ data, we tried to fit low-$T$ heat capacity data by using the relation $C_p(T) = \gamma T + \alpha T^{3/2} + \delta T^3$ (where $\gamma T$, $\alpha T^{3/2}$ and $\delta T^3$ are  the electronic,  FM and lattice contributions respectively).~\cite{gopal-book} We failed to obtain a good fit with this relation below 10 K, which reaffirms that a simple FM-like ground state is not a good picture for the studied alloy. 

\par
We have calculated the change in entropy ($\Delta S$) due to the application of $H$ (magneto-caloric effect) and it has been depicted in fig.~\ref{cp}(b). $\Delta S$ is found to be negative throughout the measurement range and it peaks around 125 K. The peak value of $\Delta S$ is about 2.8 J/kg-K for $H$ = 50 kOe and the magnitude is small compared to other Fe-doped MnNiGe alloys where magnetic and structural transitions are well separated.~\cite{liu-nc} We have also calculated the corresponding change in temperature $\Delta T$, which comes out to be -0.85 K in presence of $H$ = 50 kOe. Observation of negative $\Delta S$ is related to the shift in the MPT towards higher-$T$ in presence of external $H$. This observation is in contrast with Heusler based FSMAs, where MPT shifts towards lower $T$ on application of $H$ resulting inverse magnetocaloric effect (positive $\Delta S$).~\cite{manosa-nm,sc-jpd1} 

\section{Discussion}
Our magnetic studies on the hexagonal equiatomic alloy bring out two very important aspects which are quite important in judging the ground state magnetic properties of the studied material. Firstly, the alloy shows EBE, which clearly indicates that the ground state is not purely FM. There exists a sizable AFM or SG phase along with the majority FM fraction  which is responsible for the exchange enhanced unidirectional anisotropy in the system. The second important observation is SG like character of the ground state, which is supported by the relaxation and various memory measurements. Notably, the existence of ZFC memory rules out the possibility of the role of  superparamagnetic like clusters towards the slow dynamics of the material. The ZFC memory is an indication of cooperative freezing of the spins or spin clusters. This is only possible if there exists conflicting magnetic interactions leading to frustration. The observation of EBE warrants FM and AFM phase fractions in the sample, and magnetic frustration does take place in presence of more than one type of interactions.

\par
It is to be noted that SG and EBE are often interconnected in bulk magnetic materials.~\cite{eb-rev1} EBE is traditionally reported to occur in thin films and/or magnetic multilayeres where the magnetic characters of two layers are dissimilar. However, magnetically inhomogeneous bulk materials can also show EBE, where well defined interface is lacking. It is interesting to note that many bulk metallic spin glasses (such as AgMn or CuMn) show EBE.~\cite{eb-rev1,eb-rev2} This is due to the same magnetic inhomogeneity of the material which can give rise to complex exchange anisotropy. In such materials it is difficult to infer whether the SG phase itself gives EBE or it is the interface between SG and FM phases that is responsible for the observed effect. The present Mn$_{0.8}$Fe$_{0.2}$NiGe sample simultaneously shows SG behavior and EBE, and it is at par with the metallic spin glasses reported in the literature.~\cite{eb-rev2,sc-prb3}

\par
The ground state properties of Mn$_{0.8}$Fe$_{0.2}$NiGe has a striking similarity with the Heusler based metamagnetic shape memory alloys of general formula Ni$_2$Mn$_{1+x}$Z$_{1-x}$ (Z = Sn, Sb, In). Although both hexagonal (such as Mn$_{0.8}$Fe$_{0.2}$NiGe) and the cubic Heusler based alloys are FSMAs with the existence of first order MPT, the underlying magnetic interaction is completely different in these two groups of alloys. In the case of Ni$_2$Mn$_{1+x}$Z$_{1-x}$, excess Mn is  doped at the Z site to invoke  the martensitic instability, and the magnetic interaction between excess (at Z site) and the regular Mn atoms is of AFM in nature. SG state or EBE in these Heusler alloys occurs due to this incipient antiferromagnetism.~\cite{sc-prb3,sc-epl1} On the other hand, hexagonal alloy has only one crystallographic site for Mn and one can not have mixed magnetic bonds between Mn atoms. The likely scenario is the random occupancy of the substitutes Fe atoms at the Mn site. The parent compound MnNiGe has AFM ground state and in the undoped condition Mn-Mn interaction is AFM in nature. On substitution of Fe at the Mn site ferromagnetism is invoked possibly due to the combined effect of change in electronic and lattice properties. Due to random substitution, there may remain some Mn-rich phases, which will act as AFM islands. The interface between these AFM islands and the Fe rich FM part can give rise to conflicting magnetic interactions leading to SG like state as well as the manifestation of EBE. In addition to these similarities between Heusler and hexagonal alloys, constricted hysteresis loop is another alikeness observed for these two system of alloys.~\cite{sc-jpcs1,sc-prb3,nali1} However, heat capacity behavior of the studied hexagonal alloy is quiet different from the Heusler based FSMAs, where in the later compositions a low-$T$ linear $C_p/T$ vs. $T^2$ behavior and a shift in the MPT towards the lower $T$ with $H$ are commonly observed contrary to the former alloy.~\cite{sc-jpcm1}

\par
Fe-doped hexagonal alloy apparently resembles quite well with hole doped manganites Such as La$_{1-x}$Ca$_{x}$MnO$_3$ where both Mn$^{3+}$ and Mn$^{4+}$ exists.~\cite{pa-pr,coey-ap1} The creation of Mn$^{4+}$ favors ferromagnetism through double exchange, while Mn$^{3+}$-Mn$^{3+}$ interaction is  AFM type through superexchange. Due to random substitution there may be Mn$^{3+}$ rich AFM clusters in the otherwise Mn$^{3+}$-Mn$^{4+}$ FM background. Such electronic phase separation in managnites is found to be responsible for the glassy magnetic phase and EBE. Although the mechanism of magnetic interaction in hexagonal alloys such as Mn$_{0.8}$Fe$_{0.2}$NiGe is completely different (possibly the interaction is RKKY type, where magnetic exchange can be FM or AFM depending upon the separation of magnetic ions and the density of electronic states), a very similar FM/AFM phase separation can occur leading to the formation of SG like ground state. Notably bifurcation of ZFCH and FCH magnetization and broad peak in the ZFCH curve are very much akin with those of doped manganites.~\cite{pd-jalcom,tokura-prb2,wang-prb}

\par
EBE in hexagonal and Heusler based FSMAs resembles much with that of manganites~\cite{eb-rev2} and in both cases it points towrads the FM/AFM phase separation. However, an important difference between the intermetallic FSMAs and the manganites is that the EBE is generally observed below the spin freezing point in the former case as compared to the existence of finite EBE just below the long range magnetic order (mostly AFM type) in the later materials. It therefore points to the fact that EBE in FSMAs is closely connected with the SG state, while it is predominantly associated with the FM/AFM interfacial coupling in manganites.~\cite{salamon-eb,wu-apl1,wu-apl2,wu-scr1}    

\par
In conclusion, we present a detailed study of the magnetic ground state of Mn$_{0.8}$Fe$_{0.2}$NiGe alloy. Clear signature of SG-like ground state is evident from different magnetic measurements. Presence of ZFC memory confirms that the glassy magnetic state is associated with cooperative spin freezing. The EBE observed in the sample is closely connected with the SG fraction in the otherwise predominant FM phase of the sample.

\section{Acknowledgment}
Authors would like to thank DST, India for low temperature high magnetic field facilities at UGC-DAE Consortium for Scientific Research, Kolkata and Indore Centers.


%

\end{document}